\documentclass{emulateapj}
\usepackage{apjfonts}
\usepackage{lscape}

\slugcomment{Accepted for publication in the Astronomical Journal}

\shorttitle{Nearby stars from the LSPM Catalog. I.}
\shortauthors{S\'ebastien L\'epine}

\begin{document}

\title{Nearby stars from the LSPM-north Proper Motion
  Catalog. I. Main Sequence Dwarfs and Giants Within 33 Parsecs of the
  Sun.}

\author{S\'ebastien L\'epine}

\affil{Department of Astrophysics, Division of Physical Sciences,
American Museum of Natural History, Central Park West at 79th Street,
New York, NY 10024, USA}

\begin{abstract}
A list of 4,131 dwarfs, subgiants, and giants located, or suspected to
be located, within 33 parsecs of the Sun is presented. All the stars
are drawn from the new LSPM-north catalog of 61,976 stars with annual
proper motions larger than $0.15\arcsec$ yr$^{-1}$. Trigonometric
parallax measurements are found in the literature for 1,676 of the
stars in the sample; photometric and spectroscopic distance moduli are
found for another 783 objects. The remaining 1,672 objects are
reported here as nearby star candidates for the first
time. Photometric distance moduli are calculated for the new stars
based on the (M$_V$,V-J) relationship, calibrated with the subsample
of stars which have trigonometric parallaxes. The list of new
candidates includes 539 stars which are suspected to be within 25
parsecs of the Sun, including 63 stars estimated to be within only 15
parsecs. The current completeness of the census of nearby stars in the
northern sky is discussed in light of the new candidates presented
here. It is estimated that $\approx32\%$ ($\approx18\%$) of nuclear
burning stars within 33 parsecs (25 parsecs) of the Sun remain to be
located. The missing systems are expected to have proper motions below
the $0.15\arcsec$ yr$^{-1}$ limit of the LSPM catalog.
\end{abstract}

%   432 stars from the CNS
%   189 were new to the NLTT1 sample of Reid et al.
% 1,245 stars were in the former NLTT catalog.
%   678 are new to the LSPM-north.


%--------------------------------------------------------------------

%===%%%%%%%%%%%%%%%%%%%%%%%%%%%%%%%%%%%%%%%%%%%%%%%%%%%%%%%%%%%
\section{Introduction}

%===%
Proper motion catalogs have historically been the primary sources for
the identification of stellar systems in the vicinity of the Sun. While star
distances are ultimately determined from measurements of annual
parallactic motions, these require a substantial investment of time
and effort. Parallax programs are thus most successful when they rely
on input lists of objects already suspected to be nearby. Since a
large proper motion is an indicator of proximity, high proper motion
(HPM) stars have always been prime targets for parallax
programs. It is a historical fact that the vast majority of the stars
now known to be in the ``Solar Neighborhood'' ($d<25$ pc), and recorded
in compendiums such as the {\it Third Catalog of Nearby Stars}
\citep{GJ91} hereafter the CNS3, or the NASA {\it NStars Database}
\footnote{http://nstars.arc.nasa.gov}, have first been identified as
HPM stars. 

%===%
Perhaps the most important feature of the nearby star census is the
fact that it is dominated by low-luminosity red dwarfs, with absolute
magnitudes $9\lesssim M_V\lesssim16$. The vast majority of the ``Solar
Neighborhood'' red dwarfs are therefore not listed in the HIPPARCOS
catalog \citep{P97} which has a limiting magnitude $V\approx12$. On
the other hand, most nearby red dwarfs are expected to be detected on
the photographic plates of the large Schmidt telescope surveys
(e.g. POSS-I, POSS-II, SES, AAO) which typically reach
$V\approx20$. Because nearby red dwarfs have optical colors similar to
those of distant giants or dust-reddened stars, they are most easily
and reliably identified by their large proper motions. In fact, the
main strength of proper motion surveys is that they can locate nearby
stars independently of their physical appearance (color, luminosity).

%===%
One caveat is that a high proper motion is neither a necessary nor a
sufficient condition for a star to be nearby. Nearby stars can have
small proper motions if their motion vector is pointing toward, or
away from the Sun \citep{Getal01}. Conversely, distant stars can
have large proper motions if their transverse motion relative to the
Sun is also large, such as is typically the case for stars on Galactic
Halo orbits. Hence the dilemma: the completeness of a nearby star
census based on proper motion surveys increases as the proper motion
limit is set to smaller values, but decreasing the proper motion limit
also greatly increases the number of distant stars that make it into
the survey, which can be viewed as ``contaminants'' for parallax
programs aimed at finding the nearest stars.

%===%
The earlier proper motion catalogs of \citet{W19} and \citet{R39} had
relatively high proper motion limits ($\mu>0.3\arcsec$
yr$^{-1}$-0.5$\arcsec$ yr$^{-1}$) and together contained a little over
2500 stars, a manageable size for parallax programs. However, later
catalogs became increasingly larger, bringing the number of known high
proper motion stars to sizes difficult to manage. The catalogs by
\citet{GBT71} and \citet{GBT78} contain 11,000+ stars with $\mu>0.2$
yr$^{-1}$, while the famous ``NLTT catalog'' of \citet{L79b}, the main
reference in the field for 25 years, contains over 58,000 stars with
$\mu>0.18$ yr$^{-1}$, down to a magnitude $V=20$. These large catalogs
provide several times more targets than can be handled by existing
ground-based parallax programs \citep{Metal92,Hetal02,D02}. To
illustrate this, one need only consider the Yale catalog of
trigonometric parallaxes \citep{VLH95}, which compiles all published
ground-based parallax measurements prior to 1995, and which contains
data for only 1501 stars fainter than $V=12$. Given these limitations,
most efforts at finding nearby stars have focused on a subsample of
the NLTT, the ``LHS catalog'' \citet{L79a}, which basically lists the
4470 NLTT stars with the largest proper motions
($\mu\gtrsim0.5\arcsec$ yr$^{-1}$).

%===%
It is however possible to mine large proper motion catalogs for nearby
stars, by using secondary distance estimators such as photometric or
spectroscopic distance moduli. Photometry and spectroscopy are more
readily obtained for large samples of objects than parallax
measurements, although they provide much less reliable distance
estimates. But even approximate distance moduli can be used to trim
a large proper motion sample down to a more manageable list of probable
nearby stars. Luyten himself devoted substantial efforts to obtaining
photographic magnitudes and colors for all the stars in his catalogs,
though these were sometimes only approximate. While his $m_{pg}$,$m_r$
photographic magnitude system was used to identify some nearby objects
\citep{GJ80}, its usefulness in estimating distance moduli for the
ubiquitous red dwarfs was found to be very limited in practice
\citep{W84}. More photometric/spectroscopic data was needed, and the
follow-up VRI photometric survey of $\approx2000$ NLTT stars
with $\mu<0.5\arcsec$ yr$^{-1}$ \citep{W86,W87,W88} proved successful
in identifying 295 new Solar Neighborhood candidates. 

%===%
A major advance came with the availability of accurate infra-red
JHK$_s$ photometry from the 2 Micron All-Sky Survey (2MASS). In a
series of papers (``Meeting the Cool Neighbors. I.-IX.''), I. N. Reid
and colleagues used data from the 2MASS {\it Second Incremental
  Release} to systematically mine the NLTT for nearby stars. In the
first paper of their series, a combination of Luyten's estimated
photographic red magnitude ($m_r$) with accurate $K_s$ magnitudes from
the {\it 2MASS second incremental release} was succesfully used to
identify a subsample of candidate red dwarfs within 20 pc of the Sun
\citep{RC02}. Distance moduli were calculated based on the 2MASS
magnitudes and follow-up BVRI photometry \citep{RKC02}, and
spectroscopy \citep{CR02,Retal03} of the NLTT
candidates. Unfortunately, positions quoted in the NLTT catalog often
contain large errors \citep{GS03}; this proved a major impediment, as
2MASS counterparts of NLTT stars sometimes could not be found, or were
mismatched. These problems were mitigated by the use of the {\it
  revised NLTT} (or rNLTT) catalog of \citet{SG03}, which provide
re-estimates of NLTT positions, and through the use of additional
selection criteria (e.g. color-color cuts) to identify and reject
mismatches \citep{Retal04}.

%===%
With these successes, the nearby star census appears now to be only
limited by the NLTT catalog itself. Apart from the issues of the accuracy
of its photometry and astrometry, the NLTT has long been known to be
significantly incomplete in some parts of the sky \citep{D86}. The
nearby star census would benefit from a more complete proper motion
survey, but also from an expansion to fainter magnitudes, and
especially from an expansion to smaller proper motions, the NLTT
itself being limited to stars with $\mu>0.18\arcsec$ yr$^{-1}$.

%===%
This is why the new LSPM-north catalog, as a replacement of the old
NLTT, promises to improve significantly on nearby stars surveys. The
LSPM-north catalog \citep{LS05} is the product of a massive data
mining of the Digitized Sky Surveys (DSS) with a specialized software
(SUPERBLINK) that uses image subtraction algorithms to identify moving
and variable objects. The LSPM-north lists 61,976 stars with proper
motions $\mu>0.15$ down to V=21.0, and is estimated to be $>99\%$
complete down to $V=19$. Besides being marginally deeper, and
significantly more complete than the NLTT, the LSPM is also much more
accurate, with positions within $1\arcsec$ at the 2000.0 epoch, and
optical photographic magnitudes to $\pm0.3-0.5$mags obtained from the
USNO-B1.0 catalog of \citet{Metal03}. The LSPM-north catalog also
provides counterparts from the 2MASS {\it All Sky Point Source
  Catalog} \citep{C03} when they exist. By being deeper, more
accurate, and more complete, the LSPM-north catalog supersedes the
NLTT in all respects (though only at northern declinations for
now). The use of the 2MASS {\it All Sky Point Source Catalog} also
offers the opportunity to expand significantly upon the Reid {\it et
  al.} analysis, which was restricted to the fraction of the sky
covered by the 2MASS {\it Second Incremental Release}.

%===%
The first paper of this series presents our initial efforts at
identifying nearby stars in the LSPM-north catalog, specifically the
identification of new candidate main-sequence dwarfs within 33pc of the
Sun. Data are found in the literature for all LSPM stars that have
published trigonometric parallaxes or photometric/spectroscopic
distance moduli. A calibration of the $[M_V,V-J]$ relationship
in the $V,J$ system of the LSPM-north catalog provides the
identification of 1672 new candidate nearby stars with distances
estimated to be within 33pc of the Sun. Identification of nearby white
dwarfs from the LSPM-north catalog will be detailed in the second
paper of this series, while a subsequent paper will address the
problem of identifying nearby metal-pool subdwarfs (L\'epine, {\it in
 preparation}).

%===%
This paper is organized as follows: \S2 describes the calibration of
the $M_V,V-J$ color magnitude relationship for main sequence 
dwarfs, which is used to estimate photometric distance moduli of LSPM
stars. \S3 gives complete lists of nearby dwarfs listed in the LSPM
catalog for which there exists published trigonometric parallaxes or
photometric/spectroscopic distance moduli, and presents the selection
of new candidate stars within 33pc. \S4 analyzes the current
completeness level of the nearby star census in terms of limiting
magnitude and proper motion selection. The main results are summarized
in the conclusion (\S5).

%===%%%%%%%%%%%%%%%%%%%%%%%%%%%%%%%%%%%%%%%%%%%%%%%%%%%%%%%%%%%
\section{Stellar distance estimates}

%===%%%%%%%%%%%%%%%%%%%%%%%%%%%%%%%%%%%%%%%%%%%%%%%%%%%%%%%%%%%
\subsection{Trigonometric parallax standards}

%===%
Photometric distance moduli (m-M)$_{V,J}$ for all main sequence
LSPM-north stars are estimated based on a color-magnitude relation
in the (V,V-J) color-magnitude system used in the LSPM-north
catalog. An absolute magnitude-color calibration $M_V=M_V(V-J)$ is
calibrated with a subsample of 3,104 LSPM stars for which there exist
reliable trigonometric parallax ($\pi_{trig}$) measurements. Most
calibration stars are selected from the Hipparcos and Tycho catalog
\citep{P97}, and from the Yale catalog of trigonometric parallaxes
\citep{VLH95} \--- hereafter YPC. Calibration stars are selected based
on the following criteria: [1] the star has a measured $\pi_{trig}$
with an estimated  accuracy better than 10\%, [2] the star has
$\pi_{trig}>0.025\arcsec$, placing it  within 40pc of the Sun. [3]
the star does not have a companion within $\rho=15\arcsec$. The first
constraint ensures that absolute magnitudes can be calculated from
apparent magnitudes with an accuracy of $\pm0.1$mag. The second
constraint is used because our calibration is to be applied in
determining distances for nearby stars, and we wish to compare objects
of a given color class over a similar distance range in order to avoid
or limit biases such as dust reddening or possible systematic errors
in our USNO-B1.0 optical magnitudes. The third constraint excludes
proper motion doubles with small separations, for which V and J
magnitudes are often affected by large errors, particularly for the
faint companions.

%===%
The Hipparcos and Tycho catalog provides accurate space-based
parallaxes for 1530 LSPM stars. The Hipparcos parallaxes generally
have errors under 5 mas. Unfortunately, these are limited to
relatively bright ($V<12$) objects, which are only effective in
mapping the brighter half of the main sequence (from spectral types OB
to spectral type K). For the low-luminosity red dwarfs (spectral type
M), one has to rely on ground-based parallaxes. The YPC catalog
compiles most ground-based parallaxes published in the literature
before 1995. These, however, are not as homogeneous or accurate as the
Hipparcos parallaxes, and sometimes have errors as large as 40-50
mas. After applying the criteria above, we find reasonably accurate
YPC parallaxes for a total of 695 additional LSPM stars.

%===%
Hipparcos parallaxes can also be used for a limited number of fainter
stars that happen to be common proper motion companions of Hipparcos
objects (and are thus physically related and at the same
distance). Buy cross-correlating LSPM positions with the Hipparcos and
Tycho catalog, we find 60 LSPM stars to be common proper motion
companions of $\pi_{trig}>0.025\arcsec$ Hipparcos stars, with angular
separations $\rho>15\arcsec$ between the components. The secondaries
range in absolute magnitude from $M_V=7.8$ to $M_V=14.8$.

%===%
Despite these additions, the combined sample of Hipparcos and YPC
stars remains deficient in objects at the bottom of the main sequence
($M_V>15$). Accurate parallaxes of very low-mass M and L dwarfs have
however been published in \citet{D02} and in \citet{R03}. These papers
provide parallaxes for 13 faint LSPM stars, of which 11 have
$M_V>15$. Finally, a search of the NStars database provides reliable
parallaxes (compiled from a variety of sources) for 52 additional LSPM
stars, including 16 with $M_V>15$. The final tally is 1,951 objects to
be used as color-magnitude calibrators.

%===%%%%%%%%%%%%%%%%%%%%%%%%%%%%%%%%%%%%%%%%%%%%%%%%%%%%%%%%%%%
\subsection{Photometric distance modulus calibration}

\begin{figure} %1
\epsscale{1.25}
\plotone{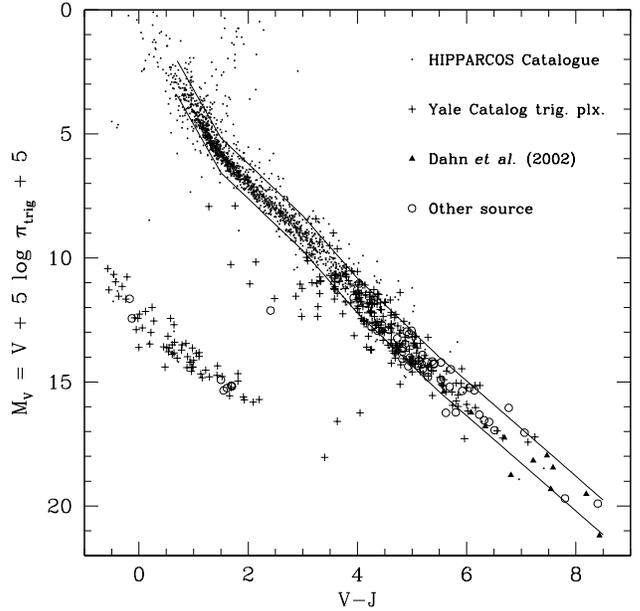}%{pxcal.eps}
\caption{Color-magnitude diagram for 1,987 stars in the LSPM catalog
  for which there are measured trigonometric parallaxes
  $\pi_{trig}>0.025\arcsec$ with 10\% or better accuracy. Different
  labels indicate the source of $\pi_{trig}$. The main sequence is
  well defined, and so is the cooling sequence of white dwarfs 
  (lower left). There is also a subgiant/giant spur discernible
  around $V-J$=2, $0<M_V<5$. Solid lines are traced $\pm0.65$mags from
  our fitted color-magnitude relationship (see text). Stars with
  $V-J>3$ falling above the lines are possibly unresolved doubles,
  while objects falling between the main sequence and white dwarf
  sequence are most probably (metal-poor) subdwarfs.} 
\end{figure}

%===%
Absolute magnitudes $M_V$ for stars in the calibration subsample are
calculated from their apparent $V$ magnitudes and observed
trigonometric parallaxes $\pi_{trig}$. These are plotted against their
$V-J$ colors in Figure 1. The main sequence is clearly visible, as is
the white dwarf sequence on the lower left of the diagram. A few stars
are found in between, and are most probably subdwarfs. The top of
the main sequence ($M_V<10$) is more densely populated than the bottom
($M_V>10$), which reflects the systematic availability of accurate
$\pi_{trig}$ for bright stars (from the Hipparcos catalog). At this
time, only limited samples of fainter stars have accurate
(ground-based) parallaxes.

%===%
For the main sequence (MS) stars, we find a monotonic relationship
between the $V-J$ color index and the absolute magnitude $M_V$. The
relationship has several inflection points. We subdivide it into 5
segments. For each, we calculate the relationship from a linear
regression of the data points (least squares fitting). We map the main
sequence as follows:
\begin{displaymath}
M_V' = 0.08 + 3.89 (V-J) \ \ \ \ \ \ \ [0.7 < V-J < 1.5]
\end{displaymath}
\begin{displaymath}
M_V' = 2.78 + 2.09 (V-J) \ \ \ \ \ \ \ [1.5 < V-J < 3.0]
\end{displaymath}
\begin{displaymath}
M_V' = 1.49 + 2.52 (V-J) \ \ \ \ \ \ \ [3.0 < V-J < 4.0]
\end{displaymath}
\begin{displaymath}
M_V' = 2.17 + 2.35 (V-J) \ \ \ \ \ \ \ [4.0 < V-J < 5.0]
\end{displaymath}
\begin{equation}
M_V' = 4.47 + 1.89 (V-J) \ \ \ \ \ \ \ [5.0 < V-J < 9.0]
\end{equation}
For any MS star, we can then calculate a photometric
distance modulus $(m-M)_{V,J}$ with:
\begin{equation}
(m-M)_{V,J} = V - M_V' ,
\end{equation}
where $M_V'=M_V'(V-J)$ follows the relationships defined
above. Photometric parallaxes $\pi_{phot}$ can then be estimated with:
\begin{equation}
\pi_{phot}= 10^{-\frac{(m-M)_{V,J}}{5}-1} .
\end{equation}

%===%%%%%%%%%%%%%%%%%%%%%%%%%%%%%%%%%%%%%%%%%%%%%%%%%%%%%%%%%%%
\subsection{Calibration accuracy}

\begin{figure} %2
\epsscale{1.25}
\plotone{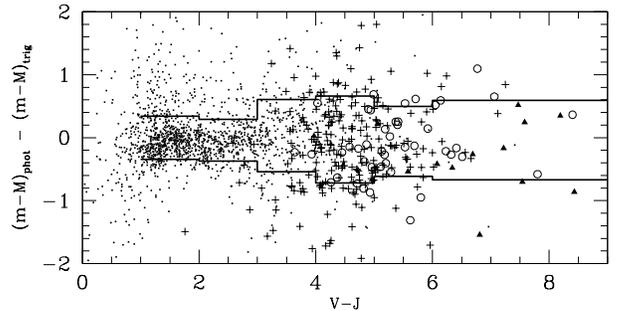}%{pxcale.eps}
\caption{Difference in the distance moduli calculated from the
  trigonometric and photometric parallaxes, for the sample of distance
  calibration stars shown in Figure 1. The solid lines show the
  running 1$\sigma$ dispersion. The dispersion is $\approx0.45$ for
  $V-J<3$ stars, which mostly consist of Hipparcos objects, for which
  both $V$ and $J$ are relatively accurate. For redder ($V-J>3$) stars,
  whose $V$ magnitudes are based on photographic plate measurements,
  the dispersion is $\approx0.65$. This indicates a mean error of up
  to $35\%$ on photometric distance estimates).}
\end{figure} 

\begin{deluxetable}{cccc}
\tablecolumns{4}
\tablewidth{0pc}
\tablecaption{Mean errors in the photometric distance moduli}
\tablehead{
\colhead{}    & \multicolumn{3}{c}{X=$(m-M)_{trig}-(m-M)_{V,J}$} \\
\cline{2-4} \\ 
\colhead{V-J} & \colhead{$\bar{X}$} & \colhead{$\sigma_X$} & \colhead{$n_X$}
}
\startdata
1.0 - 2.0 & -0.001 & 0.344 & 647 \\
2.0 - 3.0 & -0.042 & 0.331 & 372 \\
3.0 - 4.0 &  0.034 & 0.573 & 259 \\
4.0 - 5.0 & -0.027 & 0.690 & 196 \\
5.0 - 6.0 & -0.059 & 0.557 &  69 \\
6.0 - 9.0 & -0.038 & 0.631 &  32 
\enddata
\end{deluxetable}

%===%
We estimate the accuracy of our photometric distance moduli by
calculating $(m-M)_{V,J}$ for all our calibration stars, and
comparing with the trigonometric distance moduli $(m-M)_{trig}$,
where 
\begin{equation}
(m-M)_{trig} = - 5 \log{\pi_{trig}} - 5 .
\end{equation}
For each star, we calculate the difference between the trigonometric
and photometric distance moduli $(m-M)_{trig} - (m-M)_{V,J}$. These
are plotted in Figure 2 as a function of $V-J$ for our calibration
stars. We find a dispersion that increases from $\sigma\approx0.34$ mag
at $V-J<3.0$ to $\sigma=0.69$ mag at $4.0<V-J<5.0$, and down to
$\sigma=0.56$ mag at $5.0<V-J<6.0$. The values for the scatter in the
color-magnitude relationship are listed in Table 1, for given ranges
in $V-J$. These provide standard errors for the $(m-M)_{V,J}$ distance
moduli estimates.

%===%
The smaller (larger) errors on $\pi_{phot}$ for the bluer (redder)
stars is largely consistent with the expected accuracy of the $V$-band
magnitudes in the LSPM. The bluer calibration stars happen to be among
the brighter stars in the LSPM catalog ($V<12$) for which the $V$
magnitudes are quoted from the TYCHO-2 catalog \citet{H00}, and are
accurate to $\pm0.1$mag. Most of the redder calibration stars, on the
other hand, are drawn from the group of fainter LSPM stars, which have
their $V$ magnitudes estimated from photographic plate measurements,
mostly based on magnitudes from the USNO-B1.0 catalog of
\citet{Metal03} which are only accurate to $\pm0.3-0.5$mag.

%===%
One must note that at a given ($V-J$), the error in the distance
modulus is generally not equivalent to the error in the $V$
magnitude. Rather, it depends on the slope $b$ of the color-magnitude
relationship $M_V'=a+b(V-J)$. From equation 3, we see that a $\delta V$
error in $V$ propagates as $(b-1)\delta V$. Hence for $b>1$, a steeper
slope means a larger error on the photometric distance modulus for a
given $\delta V$. Hence between $4<(V-J)<5$, the $\pm0.5$ mag error
on the photographic magnitudes is expected to yield an
$\approx\pm0.68$ mag error on the distance modulus, which is consistent
with our results.

%===%
In any case, the dispersion in $(m-M)_{trig} -
(m-M)_{phot}$ cannot be attributed only to errors in the $V$-band
magnitude measurements. There is an {\em intrinsic} dispersion in the
color-magnitude relationship caused by a variety of factors such as
multiplicity, variability, and metallicity. This has been known
historically as the ``cosmic scatter''. This scatter is clear from the
$0.3$ mag dispersion observed for $(V-J)<3$ stars, whose $V$ magnitudes
are those determined from HIPPARCOS measurements, and are generally
accurate to better than $0.03$ mag. Intrinsic factors are also
probably responsible for the existence of the several outliers
($>2\sigma$ from the mean). These can be unresolved doubles,
variables, or even metal-poor subdwarfs that are naturally expected to
fall below the metal rich main sequence of the local Galactic disk
stars.

%===%%%%%%%%%%%%%%%%%%%%%%%%%%%%%%%%%%%%%%%%%%%%%%%%%%%%%%%%%%%
\subsection{Lutz-Kelker corrections}

%===%
The direct calibration of absolute magnitudes using parallax
measurements is generally susceptible to a type of systematic errors,
known as the Lutz-Kelker bias. A nice review of the topic can be found
in \citet{SS02}. This bias generally results in absolute magnitudes
being overestimated, and is dependent of the ratio of the parallax
measurement errors over the measured parallaxes $\sigma_{\pi}
\pi^{-1}$. Given that our calibrators were selected to have
$\sigma_{\pi} \pi^{-1} < 0.1$, and given the stars were selected based
on proper motion, we need to determine whether the Lutz-Kelker
is significant, and whether corrections should be applied.

%===%
Generally, the Lutz-Kelker bias occurs in a subsample of objects
with measured parallaxes $\pi_0$ and affected by measurement errors
$\sigma_{\pi}$, when the distribution of true parallaxes $\pi$ in the
underlying population decreases monotonically with $\pi$ (as when the
number of calibrators increases with distance). The measurement errors
are symmetric about $\pi_0$, but the probability distribution
$P(\pi|\pi_0)$ is not because of the asymmetry in the distribution of
$\pi$ values around $\pi_0$. Simply put, because there are more
distant stars than nearer ones, a star with measured parallax $\pi_0$
is more likely to be a distant star whose parallax is overestimated
($\pi<\pi_0$) than a nearby one whose parallax is underestimated
($\pi>\pi_0$). On average, the mean parallax of a group of stars will
thus be overestimated, which means the derived absolute magnitude will
be overestimated too. The classical case investigated by \citet{LK73}
is the one where the calibrators are uniformly distributed in space,
for which the distribution in true parallax follows
$f(\pi)\sim\pi^{-4}$. An analytical derivation indicates a bias in the
absolute magnitude calibration of +0.11 mag for $\sigma_{\pi} \pi^{-1}
= 0.10$, down to +0.02 mag for $\sigma_{\pi} \pi^{-1} = 0.05$, and
+0.01 mag for $\sigma_{\pi} \pi^{-1} = 0.025$.

%===%
All our calibrators have $\sigma_{\pi} \pi^{-1} < 0.10$, following our
selection criteria. The HIPPARCOS stars, which populate the upper half
of the main sequence ($M_V<10$), tend to have significantly more
accurate parallax measurements, with $\langle\sigma_{\pi}
\pi^{-1}\rangle = 0.036$. This value corresponds to a very small
classical Lutz-Kelker bias of only +0.014 mag. For the
calibrators that define the lower half of the main sequence
($M_V>10$), our sample has $\langle\sigma_{\pi} \pi^{-1}\rangle \simeq
0.048$, which corresponds to a bias of +0.019 mag. 

%===%
However, the above values most probably {\em overestimate} the actual
Lutz-Kelker bias in our data. The reason is that our sample of
calibrators is not drawn from the general supersample of stars in the
Solar neighborhood (whose density is uniform) but rather from a
restricted supersample that includes only stars with proper motions
above a certain value. This proper motion limit is $\mu=0.15\arcsec
yr^{-1}$ for the HIPPARCOS objects (the lower limit of the LSPM-NORTH
catalog). For the fainter calibration, it is much larger because
faint parallax targets have historically been selected among
subsamples of stars with very large proper motions (e.g. the LHS
catalog, with a proper motion limit $\mu=0.5\arcsec yr^{-1}$ ). The
effect of drawing the calibrators from a proper-motion selected
supersample tends to reduce the Lutz-Kelker bias, because the
supersample will be increasingly incomplete at larger distances (see
\S4 below), and the distribution function $f(\pi)$ will be shallower
than in the classical case. Estimates of the Lutz-Kelker bias for
shallower distribution has been investigated by e.g. \citet{H79}, who
showed that for power law distributions $f(\pi)\sim\pi^{-n}$ with
$n<4$, the magnitude of the Lutz-Kelker correction is reduced compared
with the classical case ($n=4$).

%===%
Given the uncertainty in the underlying bias associated with the
supersample (i.e. uncertainty in the form of $f(\pi)$), the
application of any specific Lutz-Kelker correction to our calibration
cannot be justified. In any case, the Lutz-Kelker bias is most
likely to be very small ($<0.02$ mag). Our photometric distances may thus
systematically underestimate the true distances of the candidate
nearby stars, but only at the $1-2\%$ level. For any given star,
this bias is negligible compared with e.g. the intrinsic ``cosmic
scatter'' of $\approx0.3$ mag, or to the $\approx0.5$ mag photometric
errors. 

%===%%%%%%%%%%%%%%%%%%%%%%%%%%%%%%%%%%%%%%%%%%%%%%%%%%%%%%%%%%%
\section{M dwarfs within 33 parsecs}

%===%
Our goal is to identify the largest possible number of dwarf stars
within the traditional confines of the Solar neighborhood
($d<25$ pc). Since our photometric distances are accurate only to
$\pm35\%$, an extension to 33pc ensures that the large majority of
25pc stars will fall into the sample. A limiting distance of 33pc is
thus adopted. The LSPM-north catalog does allow one to identify
objects at much larger distances, but the objective of this paper is
to provide a list of high priority objects of a size that can be
managed by ground-based parallax programs.

%===%%%%%%%%%%%%%%%%%%%%%%%%%%%%%%%%%%%%%%%%%%%%%%%%%%%%%%%%%%%
\subsection{Previously known objects with parallax measurements}

%===%
The LSPM-north contains thousands of new HPM stars, but lists all
previously known HPM stars as well, including those from the NLTT and
LHS catalogs. It is thus not surprising that a large number of LSPM
entries are stars that have have long been suspected (or confirmed) to
be nearby. Since the LSPM is virtually complete for all stars with
$\mu>0.15\arcsec$ yr$^{-1}$ down to a magnitude $V=19$, the vast
majority of known nearby stars from the north celestial hemisphere are
expected to have counterparts in the LSPM-north. The only exceptions
are known nearby stars with very small proper motions
($\mu<0.15\arcsec$ yr$^{-1}$) or objects with very faint optical
magnitudes ($V>20$). The latter group includes most known nearby L and
T dwarfs \citep{Ketal00,Cetal03}; these are generally too faint in the
optical bands to be detected on the POSS-I/POSS-II red plates, on
which the LSPM-north is based. In any case, most of them are not stars
but brown dwarfs.

%===%
We searched the literature for parallax measurements of LSPM sources
which would place the star within 33 pc of the Sun. Our search turned
up parallax measurements $\pi_{trig}>0.03\arcsec$ for 1676 LSPM
stars. This excludes stars whose positions in the color-magnitude
diagram suggest they are either white dwarfs or subdwarfs. The
complete list of objects is provided in Table 2; most of the stars are
main sequence objects (dwarfs) but there are a few subgiants and
giants. Parallax data are primarily from the Hipparcos and Tycho
Catalogue, including data for 1207 stars that are formal Hipparcos
objects, and for 54 more stars that are close companions or common
proper motion companions of Hipparcos objects. Another major source of
parallax data is the YPC, which provides parallaxes for 361 additional
LSPM stars. Additional $\pi_{trig}$ are found in the NStars database
for 40 more LSPM stars. Parallax data are found in \citet{D02} for 13
faint, red LSPM stars. Finally, a parallax for LSR J1835+3259 is
published in \citet{R03}.

%===%
Table 2 gives the name under which the star is generally known along
with the LSPM designation. The table quotes the LSPM position and
proper motion at the 2000.0 epoch, and gives the $V$ magnitude and
$V-J$ color, also from the LSPM. The trigonometric parallax is given
along with its estimated error, and the bibliographical source. The
distance modulus $(m-M)_{V,J}$, as estimated from the calibration
described in \S 2, is shown for comparison. Note that many stars do
not have $(m-M)_{V,J}$ estimates, because their blue color ($V-J<0.7$)
falls outside the range of of calibration system. The final column in
Table 2 gives the distance from the Sun (in parsecs) as derived from
the trigonometric parallax.

\begin{figure} %3
\epsscale{1.25}
\plotone{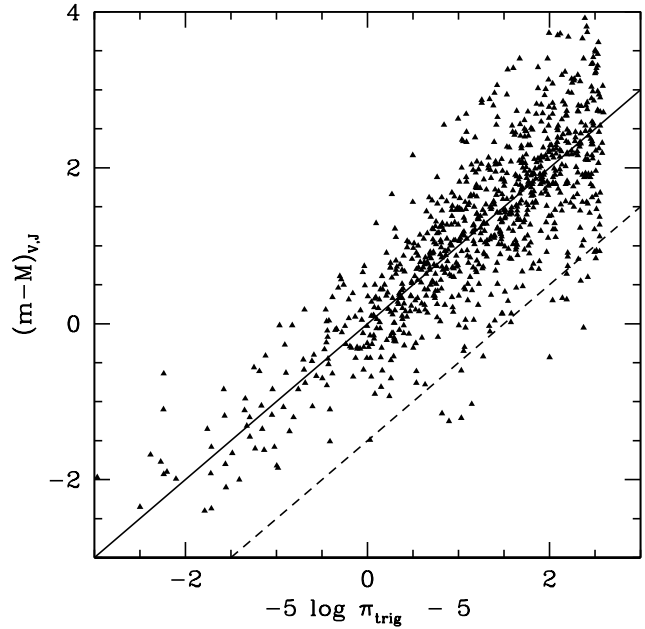}%{pxmm.eps}
\caption{Comparison between the trigonometric distance $\pi_{trig}$
  and the photometric distance modulus $(m-M)_{V,J}$ for LSPM stars
  within 33 pc that have published values of $\pi_{trig}$. The solid
  line shows the expected relationship for single stars. The
  dashed line shows the relationship expected for unresolved doubles
  of equal brightness and color. Objects close to the dashed line are
  likely to be unresolved doubles. Some stars also have 
  $(m-M)_{V,J}-(m-M)_{trig}>1$ (top of the graph); these are 
  possible metal-poor subdwarfs for which the $(m-M)_{V,J}$ overestimate 
  their actual distance.}
\end{figure} 

%===%
Figure 3 compares the derived distance modulus $(m-M)_{V,J}$ with the
distance modulus calculated from the measured trigonometric parallax
$(m-M)_{trig}=-5-5\log{\pi_{trig}}$.  The expected locus of single
stars is shown in Figure 3 (full line). The difference between the two
estimates, $(m-M)_{V,J} - (m-M)_{trig}$, has a mean value of +0.04
mag, and a dispersion 0.50 mag, after removal of the 3$\sigma$
outliers. This is consistent with our calibration of $(m-M)_{V,J}$
using a subset of these stars (see \S2 above). There appears to be a
small number of stars that have $(m-M)_{V,J} - (m-M)_{trig} \approx
-1.5$ (see Figure 3). Such an underestimate of the distance, by
photometric means, would be consistent with those stars being
unresolved doubles. The expected locus of unresolved binaries with
equal-luminosity components in the $(m-M)_{trig}$,$(m-M)_{V,J}$
diagram is drawn in Figure 3 (dashed line). There are also a few
objects whose photometric distance appear to be significantly
overestimated, with $(m-M)_{V,J} - (m-M)_{trig} > -1.5$. In all
likelihood, these objects are metal-poor subdwarfs, for which the
color-magnitude calibration of \S2 is invalid.

%===%%%%%%%%%%%%%%%%%%%%%%%%%%%%%%%%%%%%%%%%%%%%%%%%%%%%%%%%%%%
\subsection{Previously known objects with photometric/spectroscopic
  distance moduli}

%===%
There exist several hundred high proper motion stars that do not have
trigonometric parallaxes, but that have been previously cited in the
literature as candidate nearby stars. We compile a list of the main
sequence dwarfs (i.e. excluding known white dwarfs and subdwarfs) that
have been cited as candidate nearby dwarfs with estimated distance
within 33 pc, and that have a counterpart in the LSPM-north. The list,
comprising 783 stars, is provided in Table 3. Some 468 stars have
published photometric distance estimates, while another 315 have
spectroscopic distance estimates.

%===%
Some 359 stars have photometric distances quoted in the CNS3 but no
formal trigonometric parallax. Most of these stars have their distance
modulus calculated from the photometric survey of \citet{W86,W87,W88}.
Most of the other nearby star candidates are from the study of the
NLTT catalog conducted by Reid {\it et al.} We find photometric
distance estimates for 56 stars in \citet{RKC02} and for 29 more in
\citet{Retal04}. \citet{Retal03} give spectroscopic distance estimates
for 28 stars that have counterparts in the LSPM-north, and
\citet{R03} give spectroscopic distances for 169 more, and
\citet{Retal04} for 91. Spectroscopic distance estimates for high
proper motion discovered with SUPERBLINK were also obtained for 104
stars with $\mu>0.5\arcsec$ yr$^{-1}$ by \citet{LRS03}. The list
comprises 19 stars that are $d<33$pc candidates; these are included in
Table3.

%===%
There now also exists spectroscopic/photometric distance moduli for
numerous cool and ultra-cool dwarfs discovered with the 2MASS and
DENIS surveys. Photometric distance moduli are given for 52 ultra-cool
2MASS dwarfs in \citet{G00}; we find matches in that list for 15 of
the reddest LSPM stars. Spectroscopic distance moduli are calculated
for 251 late M dwarfs and L dwarfs in \citet{Cetal03}. The majority of
these objects are either in the southern sky or are too faint in the
optical to be included in the LSPM catalog, but we nevertheless find
matches to 5 additional red LSPM stars. Two nearby ultra-cool dwarfs
discovered in the DENIS infrared survey are also found among the
reddest LSPM entries: the L4 dwarf DENIS-P J104842.8+011158 has a
spectroscopic distance modulus estimated by \cite{K04}, while the
DENIS-P J031225.1+002158 has a photometric distance modulus cited in
\citet{PB01}. A photometric distance for LP 584-4 (DENIS
J000206.1+011536) is also given in \citet{PB03}. The LSPM-north
catalog also includes the very nearby star LHS 2090 ($d=6$pc), whose
photometric distance was first reported by \citet{SMJ01}. A
spectrophotometric parallax for the very high proper motion
($\mu=5.0\arcsec$ yr$^{-1}$) star SO 025300.5+165258 is given in
\citet{Tetal03}. Finally, a photometric/spectroscopic distance modulus
for LSR J0602+3910, the brightest L dwarf in the sky (V=20.1), is
given by \citet{SLRM03}.

\begin{figure} %4
\epsscale{2.40}
\plotone{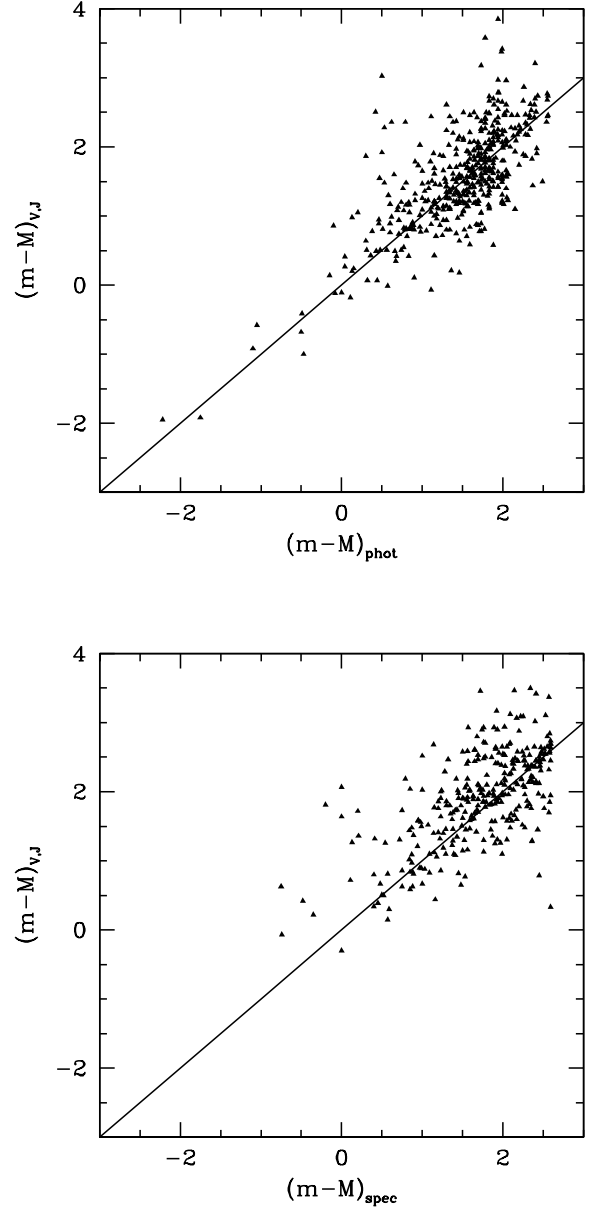}%{pxmm2.eps}
\caption{Top: comparison between photometric distances
  $\pi_{phot}$ cited in the literature and the photometric distance
  modulus $(m-M)_{V,J}$ calculated from LSPM catalog magnitudes.
  Bottom: comparison between spectroscopic $\pi_{spec}$ cited in the
  literature and the photometric distance modulus $(m-M)_{V,J}$
  calculated from LSPM catalog magnitudes. The solid line shows the
  expected relationship for single stars.}
\end{figure} 

%===%
Figure 4 compares the photometric and spectroscopic distance moduli
cited in the literature to the distance modulus $(m-M)_{V,J}$
calculated from the V magnitude and V-J color cited in the LSPM
catalog. The published $(m-M)_{phot}$ agree very well with the LSPM
distance modulus estimate. Values for $(m-M)_{V,J}-(m-M)_{phot}$ have
a mean of 0.03 mag and a dispersion of 0.43 mag, after removal of
3$\sigma$ outliers. The good match between the two distance moduli
probably only attests to the fact that the two calibrations are
essentially based on the same sample of objects with measured
$\pi_{trig}$.

%===%
On the other hand, there is a slight disagreement between the
published spectroscopic distance moduli and $(m-M)_{V,J}$, with
$(m-M)_{V,J}-(m-M)_{spec}$ having a mean value of +0.15 mag. The
excellent agreement observed between $(m-M)_{V,J}$ and both
$(m-M)_{trig}$ and $(m-M)_{phot}$ argues against any systematic error
in $(m-M)_{V,J}$. This means that the $(m-M)_{spec}$ underestimate
distances by $\approx7$\% on average. Stars with $(m-M)_{spec}$ are
overwhelmingly calibrated with the spectral-indices/absolute-magnitude
relationships defined by \citet{Retal03}. The 0.15 mag systematic error
possibly results from the limited number of stars used to calibrate
the relationships. Note that the dispersion of 0.47 mag is only
marginally larger than the dispersion on $(m-M)_{V,J}-(m-M)_{phot}$,
which suggests that $(m-M)_{spec}$ is, in principle, just about as
accurate as $(m-M)_{phot}$, provided the zero point is accurately
calibrated.

%===%%%%%%%%%%%%%%%%%%%%%%%%%%%%%%%%%%%%%%%%%%%%%%%%%%%%%%%%%%%
\subsection{New candidate nearby stars}

%===%
Using the color-magnitude calibration in \S2, it is straightforward to
identify new candidate nearby red dwarfs from the LSPM-north catalog.
Photometric distance moduli are simply estimated from $V$ and $V-J$
(Equations 1,2). The only caution is that since the relationship is
valid only for metal-rich (disk) MS stars, one must avoid applying it
to other types of objects. While the bulk of the stars in the
LSPM-north catalog are disk MS stars, the catalog also includes
significant numbers of metal-poor halo subdwarfs and white dwarfs,
along with smaller numbers of giants and subgiants.

%===%
Relative to the dwarf MS stars, giant stars are {\it
overluminous} for a given $V-J$ color. Applying the dwarf
color-magnitude relationship thus yields underestimated
distances. Giant stars are quite rare in the Solar Neighborhood, but
they can significantly contaminate samples of color-selected nearby
star candidates since they tend to be detected over a much larger
volume. Fortunately, proper motion selection filters out the majority of
more distant objects: the 0.15$\arcsec$ yr$^{-1}$ limit of the
LSPM-north catalog eliminates most disk stars beyond $\approx100$pc
(assuming disk stars to have transverse velocities
$v_t\lesssim$70km/s), and most halo stars beyond $\approx600$pc
(assuming $v_t\lesssim$400km/s). Giant stars that made it into the
LSPM-north catalog are bright enough to be HIPPARCOS objects, and
since there already is good parallaxes for them, no confusion is
possible. The more problematic objects might be the subgiants, which
are only moderately overluminous relative to the MS stars. The
LSPM-north potentially contains subgiant stars that are too faint
($V>11$) to be HIPPARCOS objects. However, these are expected to be
relatively blue ($V-J<2.5$). According to our calibration (Eq.1), main
sequence stars with $V-J<2.5$ have absolute magnitudes $M_V<8.0$,
which means they have apparent magnitude $V<10.6$ at 33pc. Since the
current goal is to identify MS stars within 33pc of the Sun, the
search can be restricted to LSPM stars with $V-J>2.5$. Any bluer
object within 33pc should have already been found and measured by
HIPPARCOS. This constraint also effectively eliminates subgiants as a
source of contaminants.

%===%
White dwarf stars, on the other hand, are {\it underluminous} for any
given $V-J$ color. As can be seen from Fig.1, the white dwarf sequence
falls $\approx$5 magnitudes below the main sequence in the standard
color-magnitude diagram. Assuming the color-magnitude relationship for
MS stars (Eq.1) were applied to a white dwarf, the object would have
its distance overestimated by a factor of 10. Restricting the search
to stars within 33pc of the Sun virtually eliminates the possibility
of a contamination of the sample with white dwarfs, because to make it
into the 33pc sample of candidate stars, the white dwarf would
actually have to be within 3.3pc of the Sun. At this time, there is no
known isolated white dwarf star within 3.3pc of the Sun. The
probability of white dwarfs ``contaminating'' the 33 pc sample is very
small.

%===%
Subdwarf stars are also {\em underluminous} in $M_V$ for a given 
$V-J$ color, although not as much as are the white dwarfs. Most nearby
subdwarfs are local member of the Galactic halo population. One way to
separate disc dwarfs from halo subdwarfs is to use an
optical-to-infrared reduced proper motion diagram \citep{SG02}. Because
the reduced proper motion is proportional to both the absolute
magnitude of a star and its transverse velocity, halo subdwarfs, with
their larger systemic velocities, fall well below the disc dwarf
sequence. But while this helps in separating halo dwarfs from disc
subdwarfs, the method only works, statistically, on large groups of
objects, and is unreliable for individual stars. In any case, halo
subdwarfs are rare in the Solar Neighborhood (about 1 star in
200); if they are found in significant numbers in proper motion
catalogs it is because of their larger mean velocity relative to the
Sun. Again, restricting the search to MS stars within 33 pc of the Sun
minimizes contamination of the nearby stars sample. The $0.15\arcsec$
yr$^{-1}$ limit includes all stars at 33 pc with transverse velocities
$v_{trans}>23.3$km s$^{-1}$, which still overwhelmingly consists of disk
stars. Increasing the distance would, however, increase the ratio of
subdwarfs to dwarfs in the census. For example, LSPM stars at 100 pc
have $v_{trans}>70.5$ km s$^{-1}$ and exclude most of the disk dwarfs,
leaving a much larger fraction of halo subdwarfs.

\begin{figure*} %5
\epsscale{1.1}
\plotone{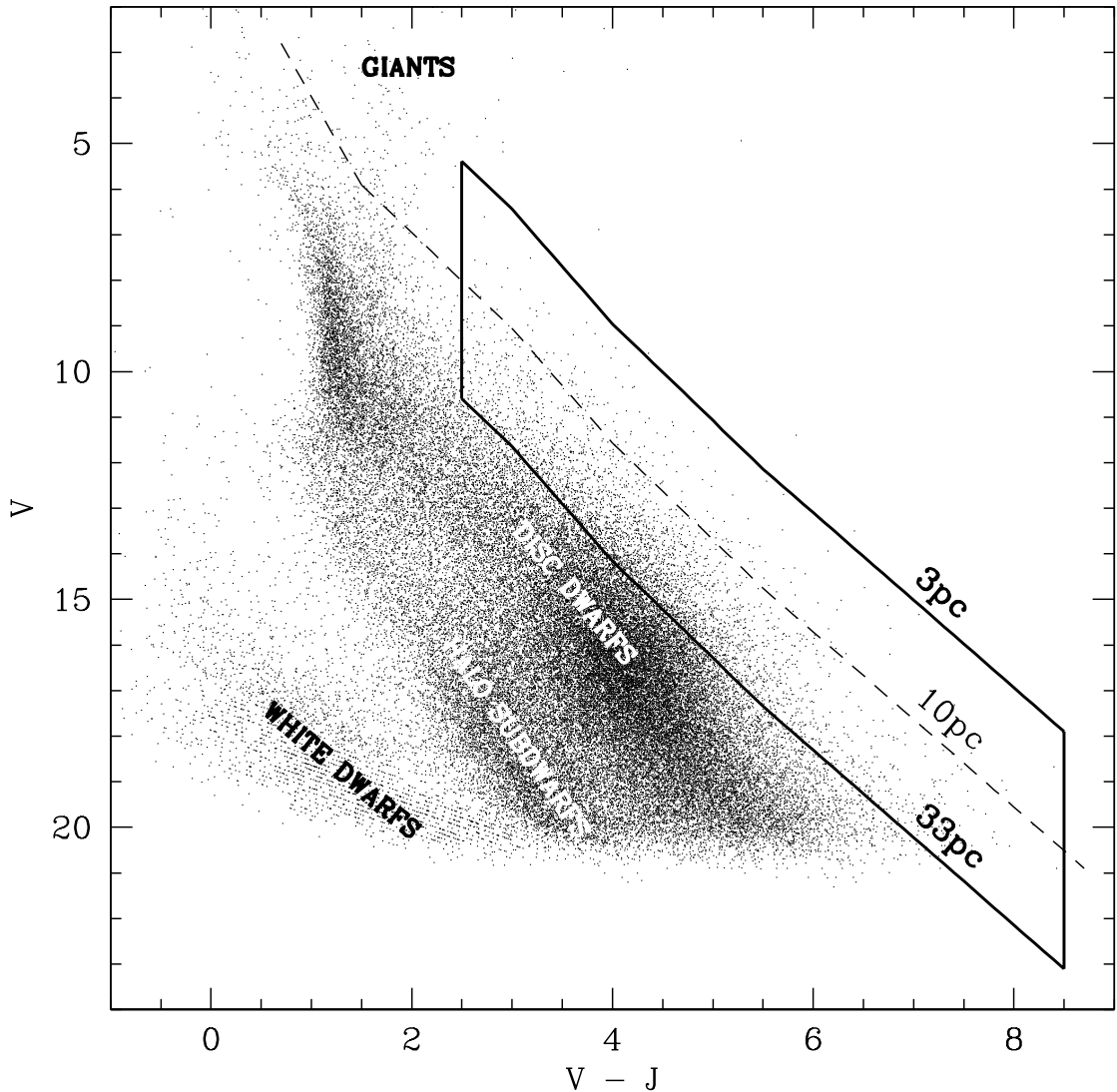}%{vvj.eps}
\caption{Selection of the nearby disc dwarfs from the high proper
  motion stars listed in the LSPM-north catalog. The color-magnitude
  relationship defined by LSPM stars with $\pi_{trig}$ measurements
  (see Figure 1) and shown above at the 10pc locus (dashed line), is
  used to select candidate nearby dwarfs between 3pc and 33pc
  (box). The typical loci of disc dwarfs, halo subdwarfs, white
  dwarfs, and giant stars are noted. The short $33$ pc distance limit
  minimizes contamination by halo subdwarfs and white dwarfs, which
  follow different color-magnitude relationships. Contamination from
  distant giant stars is avoided by selecting only objects with
  $V-J>2.5$. Note that most stars with $V<10$ already have
  $\pi_{trig}$ measurements from Hipparcos.}
\end{figure*} 

%===%
Our nearby MS stars selection box, in the $V,V-J$ diagram, is depicted
in Figure 5, overlaid on the distribution of LSPM-north stars. The
selection box marks the locus of MS stars within 3 pc and 33 pc from the
Sun. Note how the selection box avoids regions on the graph
that are mainly populated with giants, white dwarfs, and subdwarfs, as
explained above. Any star that falls in the box, and that is not a
known nearby star (i.e. it is not listed in Tables 2-3) is flagged as
a new nearby star candidate.

\begin{deluxetable*}{llrrrrrrrr}
\tabletypesize{\scriptsize}
%\rotate
\tablecolumns{10}
\tablewidth{0pc}
\tablecaption{d$<33$pc stars in the LSPM catalog: new candidate nearby
  main sequence stars\tablenotemark{1}}
\tablehead{
Name    &  LSPM \# & RA (J2000) & Decl (J2000) & $\mu$RA & $\mu$Decl  & V  & V-J & (m-M)$_{V,J}$ & dist. \\
  &  & deg. & deg. & $\arcsec$ yr$^{-1}$ & $\arcsec$ yr$^{-1}$& mag& mag & mag& pc
}
\startdata
          & LSPM J0001+0659 &  0.315899&  6.993230& -0.447& -0.081& 16.79& 5.50& 2.09$\pm$0.56& 26.2$\pm$ 8.3\\
LP 584-4  & LSPM J0002+0115 &  0.525909&  1.260014&  0.468&  0.071& 19.48& 7.31& 1.30$\pm$0.63& 18.2$\pm$ 5.7\\
          & LSPM J0005+0209 &  1.411041&  2.165013&  0.261& -0.270& 17.73& 5.74& 2.57$\pm$0.56& 32.6$\pm$10.3\\
          & LSPM J0005+4129 &  1.480111& 41.491386&  0.208&  0.012& 12.96& 3.56& 2.55$\pm$0.57& 32.3$\pm$ 9.8\\
          & LSPM J0006+0439 &  1.609240&  4.653243&  0.316& -0.253& 20.35& 7.39& 2.02$\pm$0.63& 25.4$\pm$ 8.0\\
G 217-32  & LSPM J0007+6022 &  1.927582& 60.381760&  0.342& -0.029& 13.42& 4.51& 0.82$\pm$0.69& 14.6$\pm$ 4.8\\
          & LSPM J0009+0603 &  2.323657&  6.062851&  0.094& -0.210& 15.37& 4.73& 2.30$\pm$0.69& 28.9$\pm$ 9.5\\
          & LSPM J0011+0227 &  2.785660&  2.466367&  0.097& -0.134& 15.63& 4.78& 2.46$\pm$0.69& 31.0$\pm$10.2\\
          & LSPM J0011+5736 &  2.779359& 57.614510&  0.192& -0.063& 14.66& 4.54& 2.00$\pm$0.69& 25.1$\pm$ 8.3\\
          & LSPM J0012+0206 &  3.011115&  2.106582&  0.292&  0.074& 15.40& 4.81& 2.16$\pm$0.69& 27.1$\pm$ 8.9\\
          & LSPM J0014+0213 &  3.742652&  2.218896&  0.317& -0.113& 16.81& 5.40& 2.32$\pm$0.56& 29.1$\pm$ 9.6\\
          & LSPM J0015+4344 &  3.828451& 43.742985&  0.232&  0.038& 16.69& 5.47& 2.05$\pm$0.56& 25.7$\pm$ 8.5\\
          & LSPM J0015+5829 &  3.980497& 58.494160&  0.257& -0.003& 14.86& 4.71& 1.83$\pm$0.69& 23.3$\pm$ 7.7\\
G 242-49  & LSPM J0015+7217 &  3.902023& 72.283562&  0.319&  0.185& 12.70& 3.86& 1.53$\pm$0.57& 20.3$\pm$ 6.1\\
          & LSPM J0016+3000 &  4.168521& 30.016624&  0.225&  0.028& 12.63& 3.52& 2.32$\pm$0.57& 29.1$\pm$ 8.8
\enddata
\tablenotetext{1}{The complete table is available in the electronic
  version of AJ. This table is to provides guidance to the format of the data.}
\end{deluxetable*}

%===%
Table 4 provides the complete list of new candidate nearby stars. The
lists includes 1672 stars with $8.20<V<20.63$ and $2.57<V-J<7.95$. A
total of 730 stars are objects listed in the literature or in previous
proper motion catalogs, but which have never before been formally
identified as nearby star contenders. The other 942 candidates are
high proper motion stars identified with SUPERBLINK, and first listed
in the LSPM-north catalog.

%===%%%%%%%%%%%%%%%%%%%%%%%%%%%%%%%%%%%%%%%%%%%%%%%%%%%%%%%%%%%
\subsection{New nearby star candidates of particular interest}

\subsubsection{HD 29271}

Our color-magnitude calibration puts this bright star at only
$6.3\pm1.9$ pc from the Sun. While it is a $V=8.2$ magnitude star, HD
29271 is not a HIPPARCOS star, and there is no published parallax for
it. The star is listed with a spectral type K0 in the Henry Draper
Catalog, and is also listed in the TYCHO-2 catalog with TYCHO
magnitudes $B_T=9.98$ and $V_T=8.35$. It is unclear whether this star
is an actual nearby K dwarf or a more distant K giant. With $J-H=0.53$
and $J-K_s=0.34$ (from 2MASS), this object has very unusual IR colors
which are consistent with neither a K dwarf or a K giant. Because it
has a proper motion of only 0.165$\arcsec$ yr$^{-1}$, it is more
likely to be a more distant object, although it is kept on the list of
candidate nearby stars at this time, pending clarification of its
status.

\subsubsection{G 141-36}

We estimate for this Giclas object a photometric distance for only
$7.6\pm2.4$ pc. That no follow-up observation has been performed on
this object since its identification comes as a surprise, since it
does have a large proper motion $\mu=0.447\arcsec$ yr$^{-1}$, making
it an obvious candidate for a nearby star search. This object should
be a high priority target for parallax programs.

\subsubsection{LSPM J1735+2634}

This faint ($V=19.2$) high proper motion star was first identified by
SUPERBLINK, and is one of the new stars listed in the LSPM-north
catalog. With $V-J=7.9$ and $J-K_s=1.09$, the object is most probably
a very low-mass star, of spectral type M7 or later. The photometric
distance modulus places it at a distance of $9.2\pm2.9$ pc from the
Sun.

\subsubsection{LSPM J1314+1320}

This moderately faint ($V=15.9$) star has a photometric distance
estimate of only $9.7\pm3.0$ pc. It is another one of the new high
proper motion stars found with SUPERBLINK, and also one of the
nearest. As with the one above, it should be a high priority target
for parallax programs.

\subsubsection{LSPM J1935+3746}

This $V=11.6$ star has a photometric distance estimate of only
$9.9\pm3.3$ pc. It is also a newly discovered high proper motion
star. Its moderately small proper motion (0.17$\arcsec$ yr$^{-1}$)
and location at low galactic latitudes (b=+8.33) would have made it
difficult to find in previous proper motion surveys. Even more
interesting, this star was previously found to be a counterpart of the
ROSAT point source RX J1935.4+3746 by \citet{M98} who determined a
spectral type M4Ve. The spectral type is consistent with the star
being at a distance $\approx10$ pc from the Sun.

%=...=%%%%%%%%%%%%%%%%%%%%%%%%%%%%%%%%%%%%%%%%%%%%%%%%%%%%%%%%%%%
\section{Completeness of the northern nearby dwarf census}

\begin{figure} %6
\epsscale{1.2}
\plotone{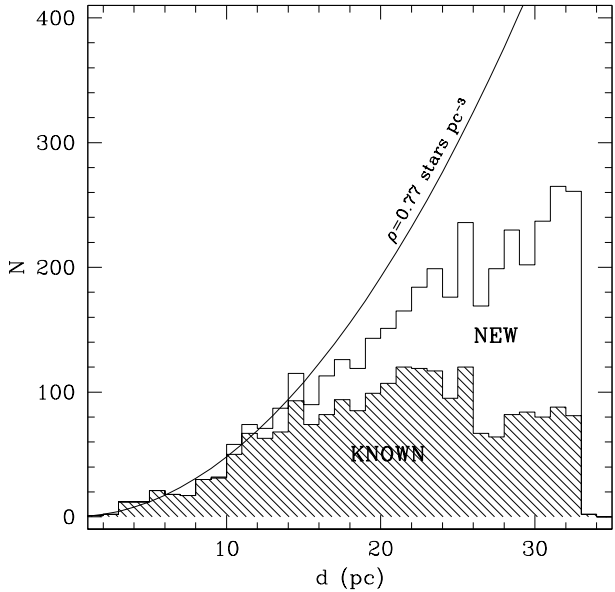}%{histo_m.eps}
\caption{Density distribution as a function of distance of northern
nearby and candidate nearby MS stars. The shaded area denotes
previously known objects. The continuous line marks the expected
distribution assuming the survey to be complete at 12pc, and a uniform
distribution of MS stars in the local volume around the Sun; the
estimated local density is 0.77 stars pc$^{-3}$. The new census
shows signs of increasing incompleteness beyond 15pc, although the
present study does make a significant contribution to the nearby star
census.}
\end{figure} 

%===%
A histogram of the distribution of nearby and candidate nearby MS
stars as a function of distance is presented in Figure 6. Stars that
were known before this study are indicated by the shaded area. The
present study makes a very significant contribution for stars beyond
20 pc; only few new additions are made within 15 pc. Assuming the current
census to be complete up to a distance of 12 pc, we estimate the local
stellar density of MS stars to be 0.77 stars pc$^{-3}$. The expected
distribution of objects as a function of distance, for a uniform
distribution of stars in the local volume, is shown in Figure 6 for
comparison.
 
%===%
The current census clearly shows signs of increasing incompleteness
beyond 15pc. This is despite the fact that the present study is
contributing a significant number of additions beyond this limit, and
especially beyond 25pc, where the new candidates account for more than
half of the known nearby objects.

\begin{figure} %7
\epsscale{1.8}
\plotone{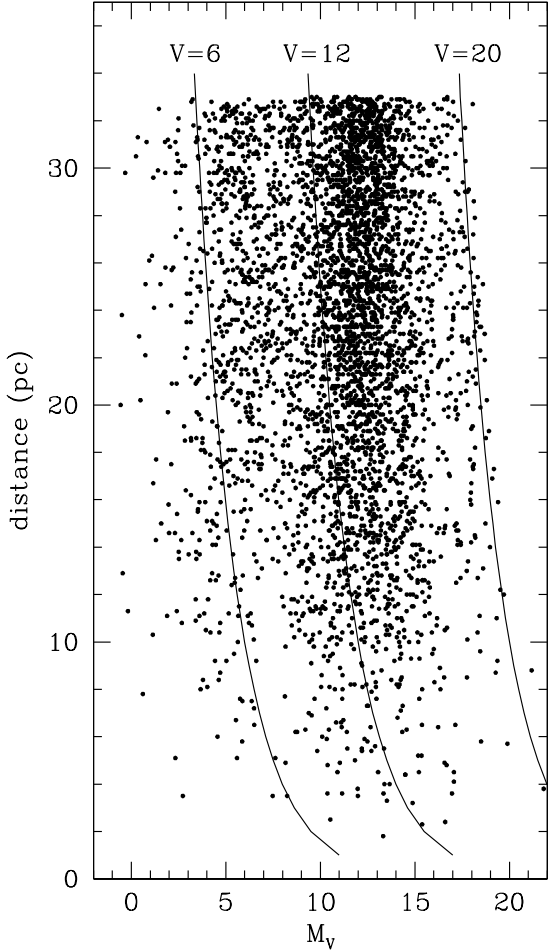}%{cmd.eps}
\caption{Distribution of nearby and candidate nearby stars as a
  function of absolute magnitude and distance. Curves indicate the
  loci of stars with apparent magnitudes $V=6$ (limit of the naked-eye
  stars), $V=12$ (limit of the HIPPARCOS catalog), and $V=20$ (limit
  of the LSPM-north catalog). The bulk of the red dwarf stars, the
  dominant constituent of the local stellar population, are in the
  range $10<M_V<15$. In the 33pc sample, most of these are beyond the
  range of HIPPARCOS, but well within the range of the LSPM-north.
}
\end{figure} 

%===%
The source of the incompleteness lies not in the limiting magnitude of
the LSPM-north catalog. The LSPM-north is estimated to 99.5\% complete
down to $V=19$, with a limiting magnitude $V=21$ \citep{LS05}. The
vast majority of the MS disk dwarfs have absolute magnitudes $M_V<16$,
and at 33 pc are all bright enough to be in the LSPM-north. Figure
7 plots the distribution of stars as a function of absolute magnitude
and distance. It is clear that the bulk of the red dwarf stars are
located between $10<M_V<15$. For stars within 33 pc, the majority are
too faint to be in the HIPPARCOS catalog, but they are well within the
magnitude limit of the LSPM-north. The 33 pc census built from the
LSPM catalog is expected to be deficient only in stars that lie at the
very bottom of the main sequence ($M_V>18$).

\begin{figure} %8
\epsscale{1.8}
\plotone{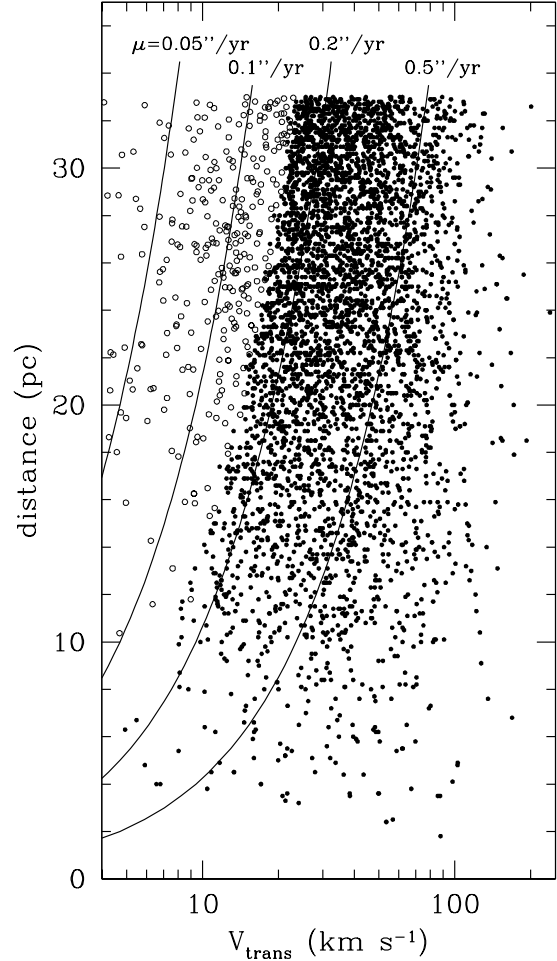}%{vit.eps}
\caption{Distribution of nearby and candidate nearby stars as a
  function of the transverse velocity and distance relative to the
  Sun. The black dots are the nearby stars that are listed
  in the LSPM-north catalog. The open circles are the known nearby
  stars that are not in the LSPM-north because their proper motion is
  below the $0.15\arcsec$ yr$^{-1}$ limit ($89\%$ are bright HIPPARCOS
  stars). This demonstrates that the proper motion limit of the LSPM
  catalog is the main source of incompleteness of the present nearby
  star census.
}
\end{figure} 

%===%
The main source of incompleteness in the 33 pc census remains the
high proper motion selection. At 33 pc, the $\mu>0.15\arcsec$ yr$^{-1}$
limit of the LSPM selects for stars with transverse velocities
$v_{trans}>23.5$ km s$^{-1}$, which excludes a significant number of disk
stars with low space motions relative to the Sun. Figure 8 plots the
distribution of nearby and candidate nearby stars as a function of
distance and transverse velocity $V_{trans}$ in km s$^{-1}$, where the
latter is calculated from the proper motion and estimated distance
($V_{trans}=4.74 \mu\ d$). Northern nearby stars that are in the LSPM
catalog are represented by black dots. The incompleteness due to
proper motion selection is clear. The very abrupt drop across the
$\mu=0.15\arcsec$ yr$^{-1}$ points to the existence of a significant
population of $d<33$pc stars with proper motions $\mu<0.15\arcsec$
yr$^{-1}$. The detection ranges of catalogs with various proper motion
limits are noted in Figure 8.

%===%
The HIPPARCOS catalog lists 265 stars in the northern sky with
$d<33$ pc and $\mu<0.15\arcsec$ yr$^{-1}$. The YPC lists 37 more
$d<33$ pc stars which are not in the LSPM-north either, because their
proper motion is too small. These 302 low proper motion nearby stars
are plotted as open circles in Figure 8. It is interesting to note
that the census of known nearby stars with {\em small} proper motions is
disproportionately biased toward HIPPARCOS ($V<12$) objects, whereas
the census of nearby stars with {\em large} proper motions is
dominated by non-HIPPARCOS ($V>12$) stars. This is consistent with the
fact that ground-based parallax programs (which the YPC is a
compilation of) have been largely dependent on lists of candidates
selected on the basis of large proper motions.

\begin{figure} %9
\epsscale{1.2}
\plotone{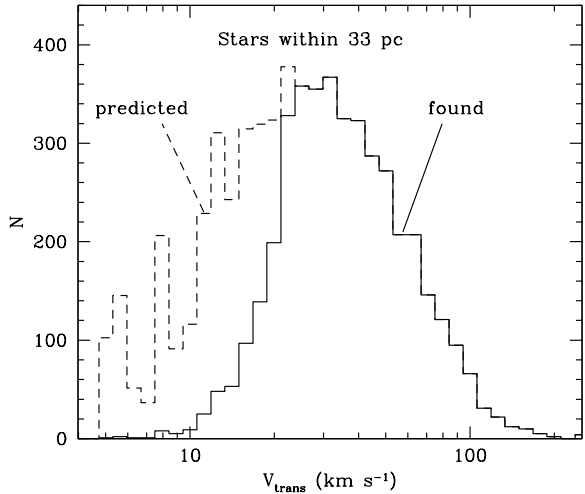}%{hvit.eps}
\caption{Full line: distribution as a function of transverse velocity
  ($v_{tran}$) of main sequence and giant stars in the current census
  of stars within 33 pc of the Sun. The census is based on a detection
  of the star in the LSPM-north proper motion catalog, which has a
  proper motion limit $\mu<0.15\arcsec$ yr$^{-1}$. Dashed line:
  predicted distribution after correcting for the fact that stars with
  low $v_{tran}$ are not detected out to the 33 pc limit because their
  proper motion is below the limit of the LSPM-north.
}
\end{figure} 

%===%
The proper motion limit results in an increasing incompleteness in the
nearby star census as a function of distance. It is possible to
estimate how many low-velocity stars are missing at larger distances
by mapping the distribution of low-velocity stars that actually get
detected at smaller distances. Figure 9 compares the distribution as a
function of transverse velocity ($v_{trans}$) for stars within 33 pc
that are actually detected by the LSPM-north, with a estimate of the
expected distribution of objects. The estimate is based on a
correction of the actual number of stars found within 33 pc at a given
$v_{trans}$ to the effective volume sampled by the LSPM-north for
stars having that value of $v_{trans}$. No correction is necessary for
velocity bins $>23.5$ km s$^{-1}$, since the $0.15\arcsec$ yr$^{-1}$
limit of the LSPM-north is sensitive to star with such velocities out
to at least 33 pc. For bins with $v_{trans}<23.5$ km s$^{-1}$, the
bins are normalized by a factor $(d_{max}/33)**3$, where
$d_{max}=v_{trans}/4.74/0.15$ is the maximum distance a star with
transverse velocity $v_{trans}$ can be detected by a proper motion
survey with $\mu=0.15\arcsec$ yr$^{-1}$ limit. The ``predicted''
velocity distribution at small $v_{trans}$ (see Figure 9) shows
fluctuations because of small number statistics, but it is clear that
at least a thousand stars must be missing from the census because of
their small proper motion. 

%===%
A simple estimate suggests that there should be $\approx6,500$ stars
within 33 pc of the Sun. The current census includes only 4434 stars,
the 4,131 ones listed in the LSPM-North, plus the 302 known nearby low
proper motion objects (see above). The current, overall completeness
of the 33pc census is thus only $\approx68\%$. This indicates that
$\approx2,000$ northern, nuclear-burning stars remain to be found
within a distance of about 33 pc, and that these have proper motions
$\mu<0.15\arcsec$ yr$^{-1}$. The situation in the 25pc volume is
slightly better, with $\approx2,800$ nuclear burning stars expected in
the northen sky, and 2,304 currently located (including 114 with
$\mu<0.15\arcsec$ yr$^{-1}$). The current completeness of the 25pc
census, the traditional extent of the ``Solar Neighborhood'', is thus
$\approx82\%$. The 500 or so missing stars are expected to have proper
motions below the limit of the LSPM-North.

\begin{figure} %10
\epsscale{1.2}
\plotone{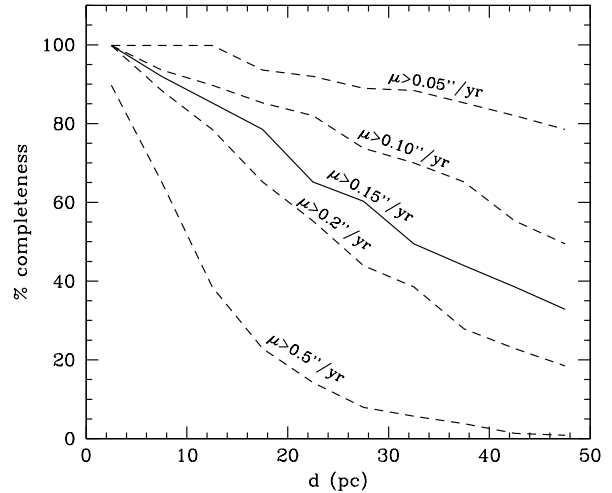}%{cvit.eps}
\caption{Predicted completeness as a function of distance of a nearby
  main sequence and giant star census based on proper motion
  catalogs. The completeness is estimated for catalogs with different $\mu$
  limits; the LSPM-north has a proper motion limit of
  $\mu<0.15\arcsec$ yr$^{-1}$ (full line). The current census of
  nearby stars based on the LSPM-north is less than 50\% complete at
  33pc. Upcoming surveys with lower proper motion limits (0.10$\arcsec$
  yr$^{-1}$, 0.05$\arcsec$ yr$^{-1}$) will increase the completeness
  significantly by adding stars with lower transverse velocities.}
\end{figure}

%===%
It is clear that much can be gained by expanding our proper motion
survey to reach smaller proper motion limits. Figure 10 plots the
expected rate of detection as a function of distance, for surveys with
lower proper motion limits $\mu=0.5\arcsec$ yr$^{-1}$, 0.2$\arcsec$
yr$^{-1}$, 0.1$\arcsec$ yr$^{-1}$, and 0.05$\arcsec$ yr$^{-1}$. These
detection rates are based on the transverse velocity distribution
shown in Figure 9. One sees that to achieve a $90\%$ detection rate at
a distance of 25 pc would require to assemble a complete survey of
stars with $\mu>0.05\arcsec$ yr$^{-1}$.

%=====%%%%%%%%%%%%%%%%%%%%%%%%%%%%%%%%%%%%%%%%%%%%%%%%%%%%%%%%%%%
\section{Conclusions}

%===%
This paper has presented a list of nearby and candidate nearby
main sequence stars, subgiants, and giants within 33 pc of the
Sun. The list is based on an analysis of the LSPM-north catalog of
stars with proper motions $\mu>0.15\arcsec$ yr$^{-1}$. The census
contains a total of 4221 stars in the half of the sky north of the
J2000 celestial equator. The list includes 1676 previously known
nearby stars with measured trigonometric parallaxes, 783 previously
suspected nearby objects, and 1762 new candidate nearby stars.

%===%
A relationship is determined between the absolute V magnitude ($M_V$)
and the $V-J$ color index. This relationship can be used to calculate
photometric distance moduli $(m-M)_{V,V-J}$ in the $V,V-J$ magnitude
system. This is convenient for estimating distances of all the stars
in the LSPM-north (for which $V$ and $J$ magnitude estimates exist) to
the extent that the relationship is applied to main
sequence(H-burning) stars of near-Solar metallicity. For single stars,
the relationship is accurate to $\pm0.5-0.6$ mag, which means that it
provides distance estimates accurate to $\pm25-30\%$. The relationship
breaks down for white dwarfs and subdwarfs, whose cases will be
addressed in upcoming papers of this series. It is found, however,
that the sample of main sequence dwarfs selected based on the
color-magnitude relationship should not be significantly contaminated
by mis-identified white dwarfs and subdwarfs, as long as the search is
limited to stars within 33 pc of the Sun.

%===%
The main purpose of the present list of candidate objects is to
provide targets for upcoming parallax programs. A fraction of the 783
previously suspected nearby MS stars are probably already on parallax
programs. The list of 1762 new candidate nearby stars should be
examined to sort out the most promising candidates to be added to the
parallax programs. Note that current distance estimates based on
photometry are only approximate, and should be used with caution.

%===%
A large spectroscopic follow-up program is under way, which will soon
provide spectral types and radial velocities for most of the new
candidate nearby stars presented here. Results will be presented in
upcoming papers in this series. While these will also yield
spectroscopic distance moduli, these are not expected to be
significantly more accurate than the photometric distance moduli
presented here. Accurate distances will ultimately come from parallax
measurements.

%===%
The census of nuclear-burning stars within 33 pc of the Sun, based on
the present study, appears to be only $\approx68\%$ for the northern
sky. The census of nuclear-burning stars within 25 pc, the traditional
confines of the so-called ``Solar neighborhood'' is estimated to be
$\approx82\%$ complete. It is revealed that the main source of
incompleteness lies in the low proper motion limit ($\mu>0.15\arcsec$
yr$^{-1}$) of the LSPM-north catalog. Expanding our survey to reach a
smaller proper motion limit appears to be the best way to
significantly increase the completeness of the nearby star census. An
extension of the SUPERBLINK survey down to a proper motion limit of
$\mu=0.10\arcsec$ yr$^{-1}$ is now being completed, and a second
extension down to $\mu=0.05\arcsec$ yr$^{-1}$ is being planned. An
analysis of the $\mu=0.05\arcsec$ yr$^{-1}$ catalog should ultimately
yield the identification of $>90\%$ of the nuclear burning stars up to
and at a distance of 25 pc.

%==%%%%%%%%%%%%%%%%%%%%%%%%%%%%%%%%%%%%%%%%%%%%%%%%%%%%%%%%%%%
%==%%%%%%%%%%%%%%%%%%%%%%%%%%%%%%%%%%%%%%%%%%%%%%%%%%%%%%%%%%%
%==%%%%%%%%%%%%%%%%%%%%%%%%%%%%%%%%%%%%%%%%%%%%%%%%%%%%%%%%%%%
\acknowledgments

{\bf Acknowledgments}

This work has been made possible by the generous support of Hilary
Lifsitz and of the American Museum of Natural History. The author is
indebted to Michael M. Shara (AMNH) and R. Michael Rich (UCLA) for
their constant support and encouragement.

%==%%%%%%%%%%%%%%%%%%%%%%%%%%%%%%%%%%%%%%%%%%%%%%%%%%%%%%%%%%%

\clearpage
\begin{landscape}
\begin{deluxetable*}{llrrrrrrrrrr}
\tabletypesize{\scriptsize}
%\rotate
\tablecolumns{15}
\tablewidth{0pc}
\tablecaption{d$<33$pc stars in the LSPM catalog: known giants,
  subgiants, and main sequence stars with trigonometric parallaxes\tablenotemark{1} }
\tablehead{
Name    &  LSPM \# & RA (J2000) & Decl (J2000) & $\mu$RA & $\mu$Decl & V & V-J & $\pi_{trig}$ & ref.\tablenotemark{2} & 
(m-M)$_{V,J}$ & dist. \\
  &  & deg. & deg. & $\arcsec$ yr$^{-1}$ & $\arcsec$ yr$^{-1}$& mag& mag & $\arcsec$ & & mag & pc
}
\startdata
 G 129-55        & LSPM J0000+1659 &  0.200457& 16.988232& -0.069& -0.305&  8.80&  1.85& 0.0318$\pm$0.0012& HIP&              & 31.4$\pm$ 1.2\\
 LHS 101         & LSPM J0002+2704 &  0.542444& 27.082132&  0.830& -0.989&  5.77&  1.07& 0.0806$\pm$0.0030& HIP&              & 12.4$\pm$ 0.5\\
 G 130-40        & LSPM J0004+2316 &  1.234669& 23.269632&  0.382& -0.007&  7.82&  1.42& 0.0391$\pm$0.0009& HIP&              & 25.6$\pm$ 0.6\\
 LHS 1014        & LSPM J0005+4547 &  1.295397& 45.786568&  0.870& -0.151& 10.05&  3.35& 0.0870$\pm$0.0014& HIP& 0.17$\pm$0.57& 11.5$\pm$ 0.2\\
 LHS 1016        & LSPM J0005+4548S&  1.420823& 45.810383&  0.839& -0.162&  8.97&  2.83& 0.0851$\pm$0.0027& HIP& 0.33$\pm$0.33& 11.8$\pm$ 0.4\\
 LHS 1017        & LSPM J0005+4548N&  1.420888& 45.812080&  0.879& -0.154&  8.83&  2.73& 0.0851$\pm$0.0027& HIP& 0.39$\pm$0.33& 11.8$\pm$ 0.4\\
 LTT 10023       & LSPM J0006+2901 &  1.653266& 29.021517&  0.381& -0.178&  6.07&  1.34& 0.0730$\pm$0.0008& HIP&              & 13.7$\pm$ 0.1\\
 G 243-13        & LSPM J0006+5826 &  1.565996& 58.436764&  0.271&  0.030&  6.36&  1.49& 0.0493$\pm$0.0010& HIP&              & 20.3$\pm$ 0.4\\
 LHS 1022        & LSPM J0007+0800 &  1.996295&  8.005391& -0.349& -0.413& 13.33&  3.94& 0.0440$\pm$0.0063& YPC& 1.96$\pm$0.57& 22.7$\pm$ 3.8\\
 V* alf And      & LSPM J0008+2905 &  2.096912& 29.090433&  0.136& -0.162&  2.05& -0.08& 0.0336$\pm$0.0007& HIP&              & 29.8$\pm$ 0.7\\
 V* V740 Cas     & LSPM J0008+6627 &  2.238476& 66.456650&  0.178&  0.002&  8.62&  1.59& 0.0313$\pm$0.0021& HIP&              & 31.9$\pm$ 2.3\\
 LHS 1027        & LSPM J0009+5908 &  2.294539& 59.149780&  0.527& -0.180&  2.27&  0.56& 0.0599$\pm$0.0006& HIP&              & 16.7$\pm$ 0.2\\
 Ross 310        & LSPM J0011+5820 &  2.838105& 58.349915&  0.235&  0.020&  9.48&  2.29& 0.0357$\pm$0.0051& HIP&              & 28.0$\pm$ 4.7\\
 CCDM J00126+214 & LSPM J0012+2142N&  3.139525& 21.713451&  0.183& -0.289& 11.91&  3.07& 0.0345$\pm$0.0117& HIP& 2.73$\pm$0.57& 29.0$\pm$14.9\\
 LP 348-42       & LSPM J0012+2142S&  3.143580& 21.706341&  0.187& -0.294& 13.65&  3.99& 0.0345$\pm$0.0117& HID& 2.16$\pm$0.57& 29.0$\pm$14.9
\enddata
\tablenotetext{1}{The complete table is available in the electronic
  version of AJ. This table is to provides guidance to the format of the data.}
\tablenotetext{2}{ References for parallaxes: HIP -- {\it Hipparcos
    and Tycho Catalogue}, HID -- {\it common proper motion companion
    of a Hipparcos star} YPC -- {\it Yale   Catalog of Trigonometric
    Parallaxes}, D02 -- {\it Dahn et al. 2002},   NST -- {\it NStars
    Database}, R03 -- {\it Reid et al. 2003b}.}
\end{deluxetable*}

\begin{deluxetable*}{llrrrrrrrrrrrrr}
\tabletypesize{\scriptsize}
%\rotate
\tablecolumns{15}
\tablewidth{0pc}
\tablecaption{d$<33$pc stars in the LSPM catalog: known main sequence stars with
  photometric/spectroscopic distance moduli\tablenotemark{1}}
\tablehead{
Name    &  LSPM \# & RA (J2000) & Decl (J2000) & $\mu$RA & $\mu$Decl & V & V-J & (m-M)$_{phot}$ & ref.\tablenotemark{2} & (m-M)$_{spec}$ & ref.\tablenotemark{3} & (m-M)$_{V,J}$ & dist. \\
  &  & deg. & deg. & $\arcsec$ yr$^{-1}$ & $\arcsec$ yr$^{-1}$& mag& mag & mag& & mag& & mag & pc
}
\startdata
LP 404-33   & LSPM J0008+2050 &  2.224675& 20.840403& -0.061& -0.255& 13.90& 5.03& 0.13$\pm$0.35& CNS&              &    & 0.20$\pm$0.56& 10.6$\pm$1.9\\
LP 191-43   & LSPM J0008+4918 &  2.229864& 49.315651&  0.347&  0.205& 16.53& 5.67&              &    & 0.99$\pm$0.17& R03& 1.50$\pm$0.56& 15.8$\pm$1.3\\
            & LSPM J0011+2259 &  2.970996& 22.984573&  0.146& -0.229& 11.98& 3.12&              &    & 1.14$\pm$0.33& R03& 2.68$\pm$0.57& 16.9$\pm$2.8\\
LHS 1037    & LSPM J0011+3303 &  2.985193& 33.054703& -0.544& -0.395& 12.98& 3.91& 1.55$\pm$0.30& R04&              &    & 1.69$\pm$0.57& 20.4$\pm$3.0\\
LP 149-35   & LSPM J0012+5059 &  3.238347& 50.988163&  0.295&  0.032& 17.24& 5.83&              &    & 1.95$\pm$0.18& R03& 1.91$\pm$0.56& 24.5$\pm$2.1\\
LP 404-66   & LSPM J0016+2003 &  4.236689& 20.065327&  0.228&  0.024& 14.11& 4.43& 1.73$\pm$0.43& CNS&              &    & 1.68$\pm$0.69& 22.2$\pm$4.8\\
LP 404-80   & LSPM J0017+2057W&  4.494271& 20.955217& -0.272& -0.384& 12.16& 3.47& 2.42$\pm$0.30& R04&              &    & 1.98$\pm$0.57& 30.5$\pm$4.5\\
LP 404-81   & LSPM J0017+2057E&  4.496541& 20.956783& -0.272& -0.384& 10.72& 2.47& 2.30$\pm$0.30& R04&              &    &              & 28.8$\pm$4.3\\
            & LSPM J0017+3028 &  4.491442& 30.469551&  0.287&  0.052& 16.28& 4.61&              &    & 2.14$\pm$0.18& R04& 3.47$\pm$0.69& 26.8$\pm$2.3\\
LP 292-66   & LSPM J0018+2748 &  4.723306& 27.813824&  0.387& -0.101& 13.80& 4.27& 1.42$\pm$0.41& CNS&              &    & 1.71$\pm$0.69& 19.2$\pm$4.0\\
LHS 1060    & LSPM J0021+1843 &  5.318830& 18.732157&  0.682& -0.084& 17.17& 5.85&              &    & 1.69$\pm$0.19& R03& 1.80$\pm$0.56& 21.8$\pm$2.0\\
LP 149-56   & LSPM J0021+4912 &  5.490994& 49.210533&  0.204& -0.031& 12.74& 3.60& 2.02$\pm$0.24& CNS&              &    & 2.23$\pm$0.57& 25.4$\pm$3.0\\
LHS 1073    & LSPM J0025+2253 &  6.335831& 22.886417& -0.241& -0.457& 14.71& 4.99& 1.00$\pm$0.19& CNS&              &    & 1.09$\pm$0.69& 15.8$\pm$1.4\\
LP 193-488  & LSPM J0026+3947 &  6.510745& 39.789875&  0.224&  0.021& 16.33& 5.34&              &    & 2.44$\pm$0.34& R03& 1.97$\pm$0.56& 30.8$\pm$5.2\\
LP 349-25   & LSPM J0027+2219 &  6.983301& 22.325680&  0.408& -0.174& 17.40& 6.79&              &    &-0.35$\pm$0.33& R03& 0.22$\pm$0.63&  8.5$\pm$1.4
\enddata
\tablenotetext{1}{The complete table is available in the electronic
  version of AJ. This table is to provides guidance to the format of the data.}
\tablenotetext{2}{ References for photometric data: CNS -- {\it 3rd
    Catalog of Nearby Stars} \citep{GJ91}, G00 -- \citet{G00},
    RC02 -- \citet{RC02}, CR02 -- \citet{CR02}, R04 -- \citet{Retal04}
    P01 -- \citet{PB01}, P03 -- \citet{PB03}, S01 -- \citet{SMJ01}, S03 --
    \citet{SLRM03}, T03 -- \citet{Tetal03}.}
\tablenotetext{3}{ References for spectroscopic data: K04 --
  \citet{K04}, LRS -- \citet{LRS03}, CR02 -- \citet{CR02}, C03 --
  \citet{C03}, R03 -- \citet{Retal03}, R04 -- \citet{Retal04}.}
\end{deluxetable*}
\clearpage
\end{landscape}


\begin{thebibliography}{}

\bibitem[Cruz \& Reid(2002)]{CR02} % paper 3: MC3 -> CR02
Cruz K.L., \& Reid I.N. 2002, \aj, 123, 2828

\bibitem[Cruz {\it et al.}(2003)]{Cetal03} % paper 5: MC5 -> C03
Cruz K.L., Reid I.N., Liebert J., Kirkpatrick J.D., \& Lowrance
P.J. 2003, \aj, 126, 2421

\bibitem[Cutri {\it et al.}(2003)]{C03}
Cutri, R. M., {\it et al.} 2003, The 2MASS All-Sky Catalog of Point
Sources University of Massachusetts and Infrared Processing and
Analysis Center (IPAC/California Institute of Technology \--- {\it
CDS-ViZier catalog number II/246})

\bibitem[Dahn {\it et al.}(2002)]{D02}
Dahn, C. C., {\it et al.} 2002, \aj, 124, 1170

\bibitem[Dawson(1986)]{D86}
Dawson, P. C. 1986, \apj, 311, 984

\bibitem[Giclas, Burnham, \& Thomas(1971)]{GBT71}
Giclas, H. L., Burnham, R., \& Thomas, N. G. 1971, Lowell proper
motion survey Northern Hemisphere. The G numbered stars. 8991 stars
fainter than magnitude 8 with motions $>$ 0".26/year, Flagstaff,
Arizona: Lowell Observatory \--- {\it CDS-ViZier catalog number
I/79})

\bibitem[Giclas, Burnham, \& Thomas(1978)]{GBT78}
Giclas, H. L., Burnham, R. Jr., \& Thomas, N. G. 1978, Lowell Proper
Motion Survey - Southern Hemisphere Catalog, Bulletin. Lowell
Observatory V. 8, P. 89 \--- {\it CDS-ViZier catalog number
I/112})

\bibitem[Gizis {\it et al.}(2000)]{G00}
Gizis, J. E., Monet, D. G., Reid, I. N., Kirkpatrick, J. D., Liebert,
J., \& Williams, R. J. 2000, \aj, 120, 1085

\bibitem[Gliese \& Jahreiss(1980)]{GJ80}
Gliese, W., \& Jahreiss, H. 1980, \aap, 85, 350

\bibitem[Gliese \& Jahreiss(1991)]{GJ91}
Gliese W., Jahreiss H. 1991, Preliminary Version of the Third
Catalogue of Nearby Stars, Astron. Rechen-Institut, Heidelberg {\it
  CDS-ViZier catalog number V/70A}

\bibitem[Gould \& Salim(2003)]{GS03}
Gould, A., \& Salim, S. 2003, \apj, 582, 1001

\bibitem[Garcia-Sanchez {\it et al.}(2001)]{Getal01}
Garcia-Sanchez, J., Weissman, P.R., Preston, R.A., Jones, D.L.,
Lestrade, J.-F., Latham D.W., Stefanik R.P., Paredes J.M. 2001, \aap,
379, 634

\bibitem[Hanson(1979)]{H79} 
Hanson, R. B. 1979, \mnras, 186, 875

\bibitem[Henry {\it et al.}(2002)]{Hetal02} 
Henry, T. J.; Walkowicz, L. M.; Barto, T. C.; \& Golimowski,
D. A. 2002, \aj, 123, 2002

\bibitem[Hog {\it et al.}(2000)]{H00} 
Hog E., Fabricius C., Makarov V.V., Urban S., Corbin T.,
Wycoff G., Bastian U., Schwekendiek P., \& Wicenec A. 2000, The
Tycho-2 Catalogue of the 2.5 Million Brightest Stars, \aap, 355, 27
({\it CDS-ViZier catalog number I/259})

\bibitem[Kendall {\it et al.}(2004)]{K04}
Kendall, T. R., Delfosse, X., Mart{\'{\i}}n, E. L., \& Forveille,
T. 2004, \aap, 416, L17

\bibitem[Kirkpatrick {\it et al.}(2000)]{Ketal00}
Kirkpatrick, J. D., Reid, I. N., Liebert, J., Gizis,
John E., Burgasser, A. J., Monet, D. G., Dahn, C. C.,
Nelson, B., Williams, R. J. 2000, \aj, 120, 447

\bibitem[L\'epine, Shara, \& Rich(2002)]{LSR02}
L\'epine, S., Shara, M. M., \& Rich, R. M. 2002, \aj, 124, 1190

\bibitem[L\'epine, Rich, \& Shara(2003)]{LRS03}
L\'epine, S., Rich, R. M., \& Shara, M. M. 2002, \aj, 125, 1598

\bibitem[L\'epine, Shara, \& Rich(2003)]{LSR03}
L\'epine, S., Shara, M. M., \& Rich, R. M. 2003, \aj, 126, 921

\bibitem[L\'epine \& Shara(2005)]{LS05}
L\'epine, S., \& Shara, M. M. 2005, \aj, 129, 1483

\bibitem[Luyten(1979a)]{L79a}
Luyten W. J. 1979a, LHS Catalogue: a catalogue of stars
with proper motions exceeding 0.5" annually, University of Minnesota,
Minneapolis ({\it CDS-ViZier catalog number I/87B})

\bibitem[Luyten(1979b)]{L79b}
Luyten W. J. 1979b, New Luyten Catalogue of stars with
proper motions larger than two tenths of an arcsecond (NLTT),
University of Minnesota, Minneapolis ({\it CDS-ViZier catalog number
I/98A})

\bibitem[Lutz \& Kelker(1973)]{LK73}
Lutz, T. E., \& Kelker, D. H. 1973, \pasp, 85, 573

\bibitem[Monet {\it et al.}(1992)]{Metal92}
Monet, David G., Dahn, Conard C., Vrba, Frederick J., Harris, Hugh C.,
Pier, Jeffrey R., Luginbuhl, Christian B., \& Ables, Harold D. 1992,
\aj, 103, 638

\bibitem[Monet {\it et al.}(2003)]{Metal03}
Monet, D. G., {\it et al.} 2003, \aj, 125, 984, (The USNO-B1 catalog
\--- {\it CDS-ViZier catalog number I/284})

\bibitem[Motch {\it et al.}(1998)]{M98}
Motch, C., Guillout, P., Haberl, F., Krautter, J., Pakull, M.W.,
Pietsch, W., Reinsch, K., Voges, W., \& Zickgraf, F.-J. 1998, \aaps,
132, 341

\bibitem[Perryman(1997)]{P97}
Perryman, M. A. C. 1997, The Hipparcos and Tycho
catalogues. Astrometric and photometric star catalogues derived from
the ESA Hipparcos Space Astrometry Mission, Publisher: Noordwijk,
Netherlands: ESA Publication ({\it CDS-ViZier catalog number I/239})

\bibitem[Phan-Bao {\it et al.}(2001)]{PB01}
Phan-Bao, N., Guibert, J., Crifo, F., Delfosse, X., Forveille, T.,
Borsenberger, J., Epchtein, N., Fouque, P., \& Simon, G. 2001, \aap, 380,
590

\bibitem[Phan-Bao {\it et al.}(2003)]{PB03}
Phan-Bao, N., Crifo, F., Delfosse, X., Forveille, T., Guibert, J., 
Borsenberger, J., Epchtein, N., Fouque, P., Simon, G., \& Vetois, J. 2003,
\aap, 401, 959

\bibitem[Reid \& Cruz(2002)]{RC02} %paper1:  MC1 -> RC02
Reid, I. N., \& Cruz, K. L. 2002, \aj, 123, 2806

\bibitem[Reid, Kilkenny \& Cruz(2002)]{RKC02} %paper2
Reid, I. N., Kilkenny, D., \& Cruz, K. L. 2002, \aj, 123, 2822

\bibitem[Reid {\it et al.}(2003a)]{Retal03} %paper7: MC7 -> R03	
Reid, I. N., Cruz, K. L., Allen, P., Mungall, F., Kilkenny, D.,
Liebert, J., Hawley, S. L., Fraser, O. J., Covey, K. R., \& Lowrance,
P. 2003a, \aj, 126, 3007

\bibitem[Reid {\it et al.}(2003b)]{R03} %paper4
Reid, I. N., Cruz, K. L., Laurie, S. P., Liebert, J., Dahn, C. C.,
Harris, H. C., Guetter, H. H., Stone, R. C., Canzian, B., Luginbuhl,
C. B., Levine, S. E., Monet, A. K. B., Monet, D. G. 2003b, \aj, 125,
354

\bibitem[Reid {\it et al.}(2004)]{Retal04} %paper8 MC8: R04	
Reid, I. N., Cruz, K. L., Allen, P., Mungall, F., Kilkenny, D.,
Liebert, J., Hawley, S. L., Fraser, O. J., Covey, K. R., Lowrance, P.,
Kirkpatrick, J. D., \& Burgasser, A. J. 2004, \aj, 128, 463

\bibitem[Ross(1939)]{R39}
Ross, F. E. 1939, \aj, 48, 1118

\bibitem[Salim \& Gould(2002)]{SG02}
Salim, S., \& Gould, A. 2002, \apj, 575, L83

\bibitem[Salim \& Gould(2003)]{SG03}
Salim, S., \& Gould, A. 2003, \apj, 582, 1011

\bibitem[Salim {\it et al.}(2003)]{SLRM03}
Salim, S., L\'epine, S., Rich, R. M., \& Shara, M. M. 2003, \apjl,
586, L149

\bibitem[Sandage \& Saha (2002)]{SS02}
Sandage, A., \& Saha, A. 2002, \aj, 123, 2047

\bibitem[Scholz, Meusinger, \& Jahreiss(2001)]{SMJ01}
Scholz R.-D., Meusinger H., \& Jahreiss H. 2001, \aap, 374, L12

\bibitem[Teegarden {\it et al.}(2003)]{Tetal03}
Teegarden B.J., Pravdo S.H., Hicks M., Lawrence K., Shaklan S.B., 
Covey K., Fraser O., Hawley S.L., McGlynn T., \& Reid I.N. 2003, 
\apjl, 589, L51

\bibitem[van Altena, Lee, \& Hoffleit(1995)]{VLH95}
Van Altena W. F., Lee J. T., Hoffleit E. D. 1995, The General
Catalogue of Trigonometric Stellar Parallaxes, Fourth Edition, Yale
University Observatory 1995 ({\it CDS-ViZier catalog number I/238A})

\bibitem[Weis(1984)]{W84}
Weis, E. W. 1984, \apjs, 55, 289

\bibitem[Weis(1986)]{W86}
Weis, E. W. 1986, \aj, 91, 626

\bibitem[Weis(1987)]{W87}
Weis, E. W. 1987, \aj, 93, 451

\bibitem[Weis(1988)]{W88}
Weis, E. W. 1988, \aj, 96, 1710

\bibitem[Wolf(1919)]{W19}
Wolf, M. 1919, Katalog von 1053 starker bewegten Fixternen,
Veroeff. des Badischen Sternw. Heidelberg, 7, 10 

\end{thebibliography}
\end{document}